\pdfoutput=1

\documentclass[11pt]{article}

\usepackage{acl}

\usepackage{times}
\usepackage{latexsym}

\usepackage[T1]{fontenc}

\usepackage[utf8]{inputenc}

\usepackage{microtype}

\usepackage{graphicx}
\usepackage{multirow}
\usepackage{amsmath}
\usepackage{multirow}
\usepackage{arydshln} 
\usepackage{float} 
\usepackage{caption}
\usepackage{subcaption}

\title{Rescue Conversations from Dead-ends: Efficient Exploration for \\ Task-oriented Dialogue Policy Optimization}

 \author{Yangyang Zhao$^{123}$\thanks{\ \ Corresponding author}, Zhenyu Wang$^{3}$, Mehdi Dastani$^{2}$ \and Shihan Wang$^{2}$ \\
	$^1$Changsha University of Technology \\ $^2$Utrecht University  \\ $^3$South China University of Technology  \\  \texttt{msyyz@mail.scut.edu.cn; wangzy@scut.edu.cn;}\\\texttt{m.m.dastani@uu.nl; s.wang2@uu.nl}}

\begin{document}
\maketitle
\begin{abstract}
Training a dialogue policy using deep reinforcement learning requires a lot of exploration of the environment. The amount of wasted invalid exploration makes their learning inefficient. In this paper, we find and define an important reason for the invalid exploration: dead-ends. When a conversation enters a dead-end state, regardless of the actions taken afterward, it will continue in a dead-end trajectory until the agent reaches a termination state or maximum turn. We propose a dead-end resurrection (DDR) algorithm that detects the initial dead-end state in a timely and efficient manner and provides a rescue action to guide and correct the exploration direction. To prevent dialogue policies from repeatedly making the same mistake, DDR also performs dialogue data augmentation by adding relevant experiences containing dead-end states. We first validate the dead-end detection reliability and then demonstrate the effectiveness and generality of the method by reporting experimental results on several dialogue datasets from different domains.
\end{abstract}

\section{Introduction}

Task-oriented dialogue (ToD) systems are designed to help users accomplish a pre-defined goal, such as booking a restaurant or movie tickets \cite{peng2018deep, ZhangLGC19, wu2021gaussian}. Typically, ToD systems are built on a structured database (DB) containing many task-related entries for retrieving information \cite{DBLP:conf/lrec/PolifroniW06, LiYQY21}. Every entry of DB is represented in terms of a set of attributes (refers to slot) and their values. For example, in the movie domain, slots for the movie entity include \textit{movie name, theater, and date}. In order to interact with users, ToD systems require a dialogue policy that determines how the system should reply to the user's input \cite{chen2017survey, WangPW20, kwan2022survey}. A well-performing dialogue policy is expected to capture as much user constraint information as possible in a few dialogue turns and find matching entries in the DB to accomplish user goals \cite{geishauser2021does}. Thus, the dialogue policies are often modeled as a Partially Observable Markov Decision Process (POMDP) \cite{young2013pomdp}
\footnote{Theoretically, if a Markov Decision Process (MDP) \cite{puterman1990markov} corresponding to an environment is fully specified, then its (optimal) value function can be computed without any empirical interaction with the actual environment.}
and optimized using Deep Reinforcement Learning (DRL) \cite{su2016line, peng2018adversarial, TakanobuLH20}.

\begin{figure*}[t]
	\centering
	\subcaptionbox{Examples of an invalid exploration dialogue trajectory \label{maina}} {\includegraphics[width=0.65\columnwidth,trim=50 10 450 0,clip]{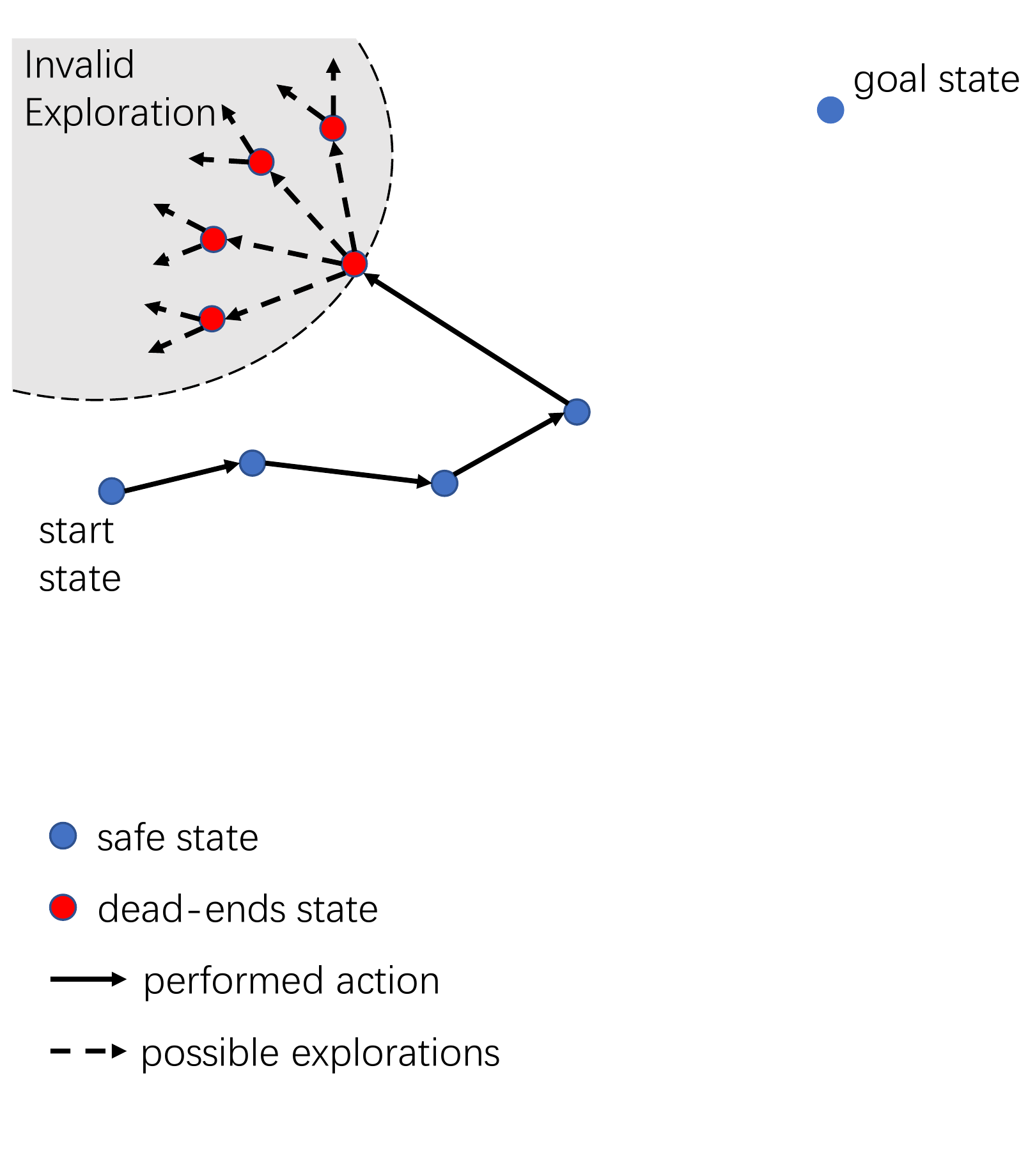}}
	\subcaptionbox{Illustration of DDR with information gain exploration guide (DDR-IG) \label{mainb}} {\includegraphics[width=0.65\columnwidth,trim=50 10 450 0,clip]{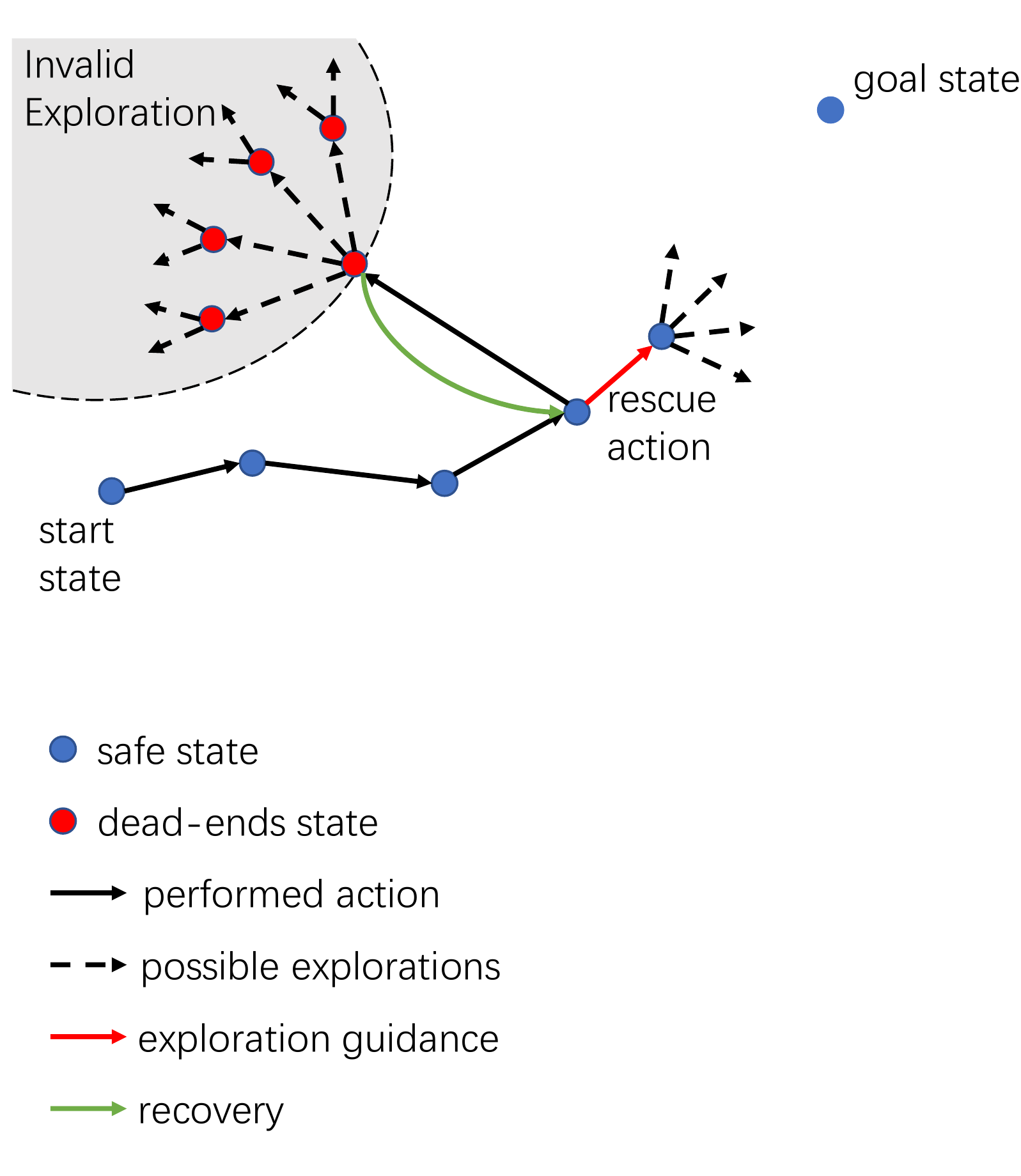}}
	\subcaptionbox{Illustration of DDR with self-resimulation exploration guide  (DDR-SE) \label{mainc}}{\includegraphics[width=0.65\columnwidth,trim=50 10 450 0,clip]{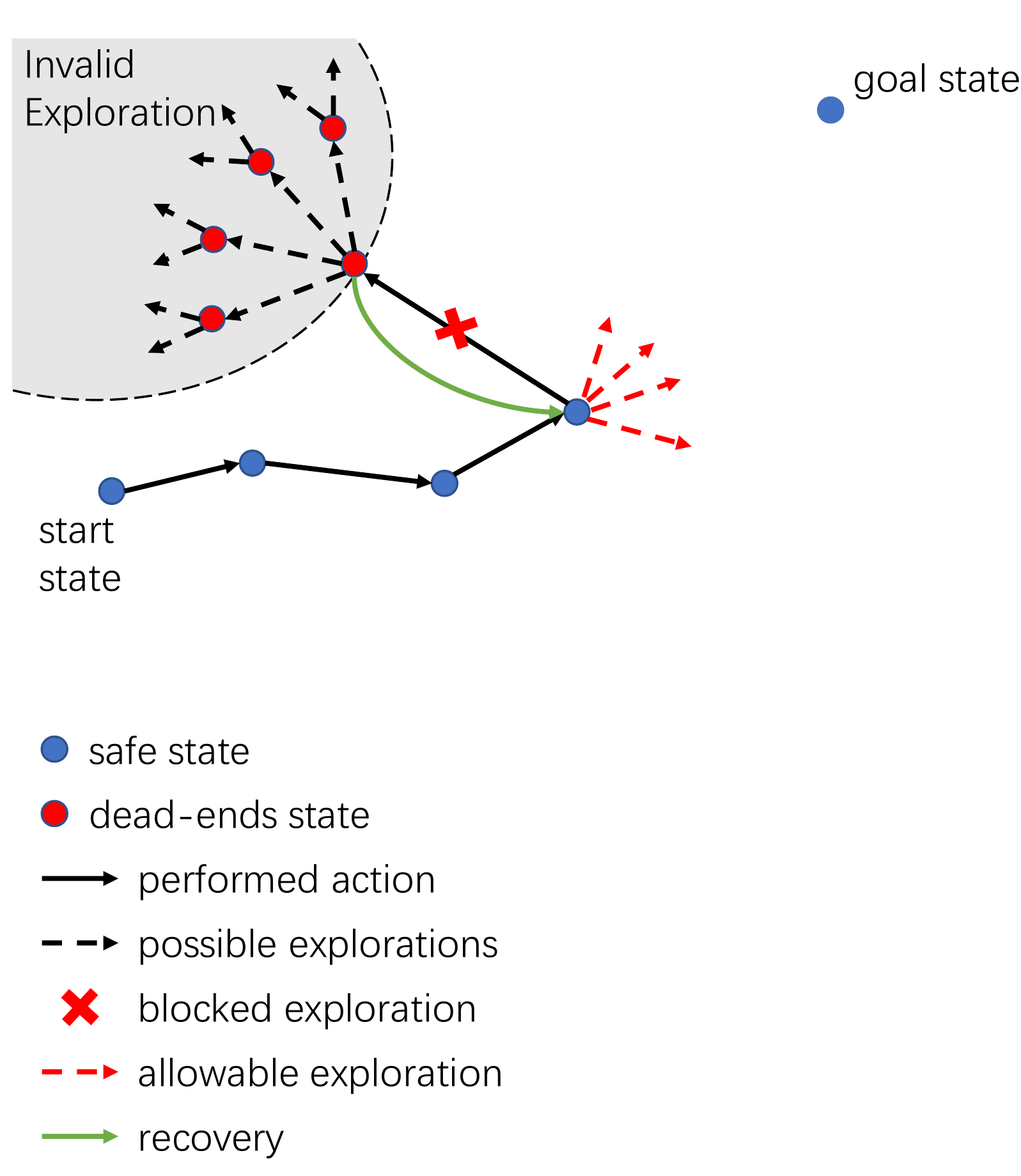}}
	\caption{Illustration of invalid exploration  (a), and our two dead-end resurrection (DDR) algorithms (b and c).}
	\label{fig:main}
\end{figure*}

However, in many real-world dialogue scenarios, transition probabilities or rewarding transitions are not known in advance, and dialogue agents usually require extensive exploration of the environment \cite{li2016user, kwan2022survey}. Some of these explorations are valid, and some are not, and the more invalid explorations there are the less effective learning of dialogue policies become \cite{wu2019tam, wu2021clipping}. 
In many dialogues, a long dialogue segment is an invalid exploration due to the effect of dead-ends \footnote{Our definition of the dead-end is inspired by \citet{DBLP:conf/icml/FatemiSSK19}, but our definition and detection are not the same due to the different research domains between the game domain and the dialogue domain (see Section~\ref{sec:definition} for a formal definition).}. As shown in Fig.~\ref{maina}, a dialogue agent can enter a dead-end due to the accumulation of errors in the upstream module or its own decision failures. After reaching the dead-end state, the agent will always end up at a failed terminal state after some steps, regardless of whatever action it chooses. Such dialogue segments that start from a dead-end state cause invalid explorations and, thus less effective learning of dialogue policies. Furthermore, such invalid explorations inherently produce a tremendously large number of training steps without getting useful information to achieve the goal. In other words, it makes the dialogue longer by having a trajectory of unnecessary interactions, which makes it more difficult for the dialogue agent to learn which responses are bad among a lengthy dialogue.

This paper proposes a dead-end resurrection (DDR) algorithm that detects the initial dead-end state in a timely and efficient manner and provides a rescue action to guide and correct the exploration direction. Our DDR method can effectively mitigate the adverse effects of dead-ends, improve exploration efficiency, and provide more effective and diverse samples for training. 

Specifically, we define an effective and reliable dead-end detection criterion for ToD systems that work with DB. The current state is considered the initial dead-end state if the matching entries in DB go from available to none at a certain moment. 
Once a certain dead-end is detected, a rescue module in the DDR is activated, which transfers the current dead-end state to the previous safe state and provides rescue guidance to guide and correct the direction of exploration.
Our DDR approach considers two types of rescue guidance. As shown in Fig.~\ref{mainb}, the first rescue guidance directly provides an explicit rescue action with maximum information gain. The second rescue guidance is shown in Fig.~\ref{mainc}. It blocks the wrong path for the dialogue agent and allows the agent to explore the diverse path. We report a number of experiments where we investigated the advantages and applicability of these two rescue guidances. 

In addition, to prevent dialogue policies from continuously making the same mistakes, a warning experience, including the dead-end state and a negative penalty, is added to the experience pool for dialogue data augmentation (DDA). The effectiveness of our DDR algorithm is verified on four public datasets while remaining robust against noisy environments. Importantly, our approach is a model-agnostic algorithm with sufficient generality to be applied to different RL-based dialogue policy methods. In summary, our contributions are as follows:
\begin{itemize}
	\item  Define the concept of the dead-end for dialogue policy exploration and demonstrate their widespread existence and impact on the learning efficiency of dialogue policy.
	\item Proposed an efficient dead-end detection criterion and verified its reliability.
	\item Proposed dead-end resurrection (DDR) algorithm that effectively rescues conversations from dead-ends and provides a rescue action to guide and correct the exploration direction.
	\item Validated the effectiveness and generalizability of the DDR approaches on several datasets.
\end{itemize}

\section{Related Work}

\begin{figure}[!t]
	\centering
	\includegraphics[scale=0.38,trim=180 0 100 0,clip]{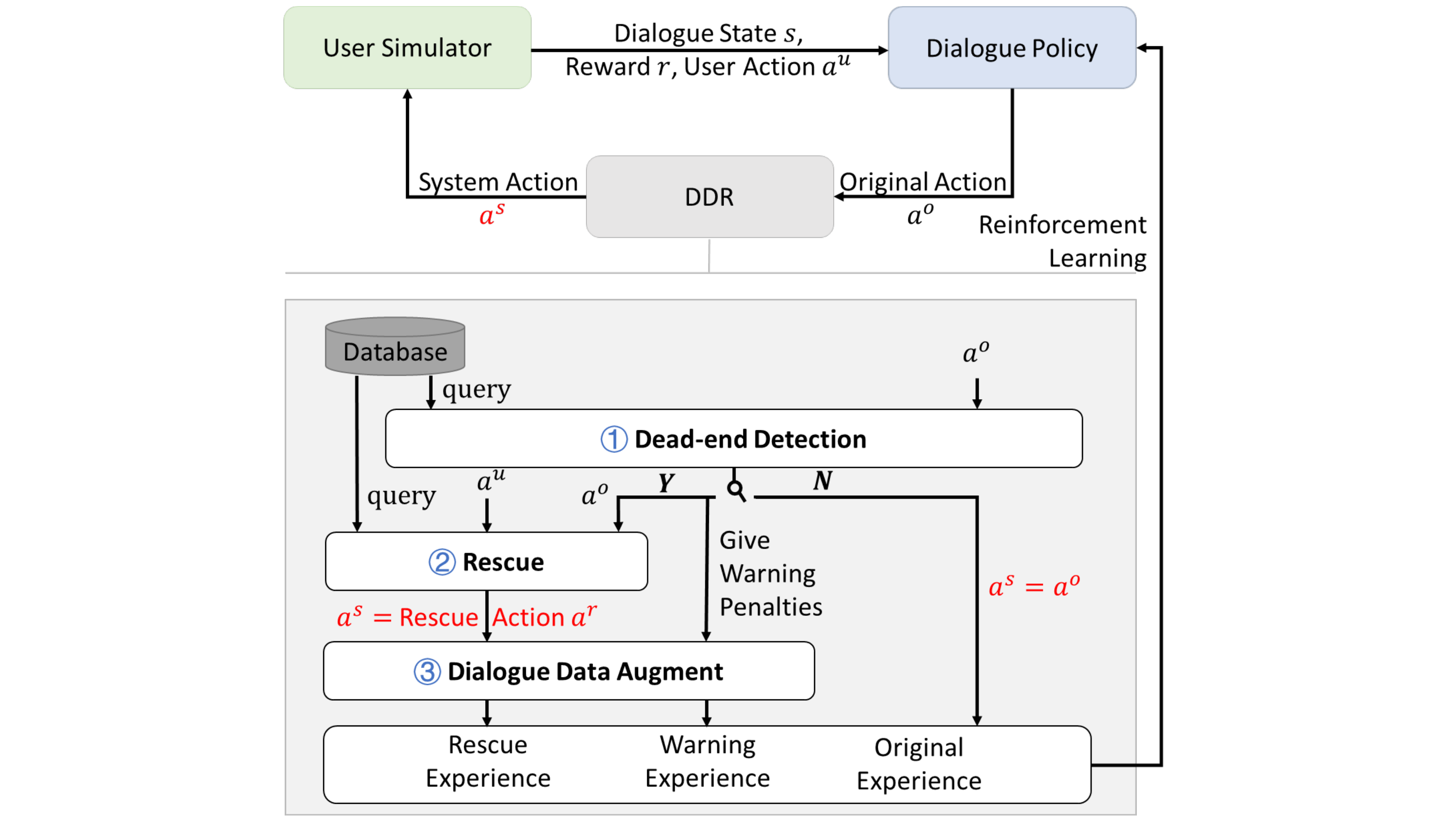}
	\caption{Illustration for dialogue policy optimization using proposed DDR algorithm.}
	\label{fig: method}
\end{figure}
Our work connects to several research attempts to apply RL to learn dialogue policies for task-oriented dialogue systems. We touch on the research works aiming to make RL learning efficient by improving sample efficiency, \uppercase\expandafter{\romannumeral1}) developing effective exploration strategies to generate more effective samples, and \uppercase\expandafter{\romannumeral2}) directly manipulating generated training samples. Since both our DDR approach and the Type \uppercase\expandafter{\romannumeral2} approaches have the potential to combine with the above exploration strategies to produce a more effective learning process, in this paper, we compare with the Type \uppercase\expandafter{\romannumeral2} approaches based on the same exploration strategy \uppercase\expandafter{\romannumeral1}.

Type \uppercase\expandafter{\romannumeral1}: Research in the field of reinforcement learning has developed many studies on general and effective exploration strategies to address the issue of sample efficiency. Such as the most basic exploration method $\epsilon$-greedy \cite{sutton1995generalization}, Boltzmann exploration \cite{kaelbling1996reinforcement,szepesvari2010algorithms}, Upper confidence bounds  \cite{lai1985asymptotically}, Gaussian exploration \cite{lillicrap2015continuous}, exponential gradient $\epsilon$-greedy \cite{li2010exploitation}, etc. More recently, intrinsic motivations have been introduced to encourage exploration \cite{schmidhuber1987evolutionary, singh2010intrinsically, bellemare2016unifying}—for instance, curiosity-based exploration \cite{pathak2017curiosity}, uncertainty reduction \cite{lopes2012exploration, houthooft2016vime}, Tompson sampling \cite{chapelle2011empirical, osband2016deep}, the Bayes-by-Backprop Q-network \cite{lipton2018bbq}. 
Due to its popularity and simplicity, this paper defaults to using the $\epsilon$-greedy exploration strategy for exploration unless otherwise noted.

Type \uppercase\expandafter{\romannumeral2}: Some studies in the field of task-oriented dialogue policies improve sample efficiency by manipulating the training samples. \citet{chen2017line,chen2017agent}  introduced companion learning to provide the necessary guidance for dialogue policies. However, such an approach transfers the training burden of dialogue policies to the cost of designing the teacher model that provides guidance \cite{zhao2021efficient}. Hindsight experience replay (HER) is proposed to extract partial success segments from failure dialogues to synthesize successful artificial dialogues for DDA \cite{andrychowicz2017hindsight, lu2019goal}. However, the quality of the synthetic experience highly affects the effectiveness of HER \cite{cao2020adaptive}. Trainable-Action-Mask (TAM) learns action masks from the data to block useless actions to generate more effective samples \cite{wu2019tam}. 
Loop-Clipping Policy Optimisation (LCPO) clips useless trajectories to improve sample efficiency \cite{wu2021clipping}. But both the experiences generated by LCPO and HER are non-exploratory and non-authentic. Instead, our approach allows exploration to generate diverse, high-quality, successful dialogues without artificial synthesis and parameter tuning. Essentially, they only process the dialogue fragments after their end, while our DDR detects and fixes the dialogue during the conversation.

\section{Method}

As illustrated in Fig.~\ref{fig: method}, DDR consists of three main stages: 1) \textit{Dead-end Detection} process automatically detects whether the current conversation with the original action $a^o$ leads to a dead-end. If so, 2) \textit{Rescue} process transfers the current dead-end state back to the previous safe state and provides a rescue action $a^r$ as system action $a^s$ for the response. And 3) \textit{Dialogue Data Augment} process is performed by adding a warning experience containing this dead-end state and warning penalties into the experience buffer to prevent the dialogue policy from repeatedly making the same mistake. Otherwise, the original action $a^o$ is transferred directly to the user simulator as a response $a^s$ without \textit{Rescue}  and \textit{Dialogue Data Augment} stages. These three stages are described in detail in the following subsections.

\subsection{Dead-end Definition and Detection}
\label{sec:definition}

This section introduces the definition and detection of dead-end in the dialogue policy task.

\newtheorem{definition}{Definition}
\begin{definition}
	(Dead-end). A state $s$ is called a dead-end state if all trajectories starting from this state fail with probability one after several turns. 
\end{definition}

If a state is a dead-end state, all possible states following this state are also dead-end states. We use dead-end to denote the trajectory from the initial dead-end state to the termination state of this conversation trajectory.

When a conversation enters a dead-end state $s$, regardless of actions taken afterward, it will continue in a dead-end trajectory until the agent reaches a termination state or maximum turn. To this end, we need to detect the initial state of the dead-end (referred to as dead-end detection) to avoid invalid exploration.

Note that the purpose of the dialogue agent is to obtain constraint information through interactions and find entries from the database that satisfy all the constraints to accomplish the user goal. During the conversation, the number of matching entries changes as the user provides more constraints guided by the dialogue agent. The number of matching entries that satisfy the user's constraints, denoted as $n$, at a particular stage of the dialogue is an important signal to determine whether a dialogue action will lead to a conversation failure\footnote{In dialogue tasks, $n$ is a feature of the dialogue state, and it is easy to obtain in the ToD system.}. When the value of $n$ decreases to 0, then the current conversation fails into a dead-end. 
In this way, our detection way can potentially identify dead-ends in time during the conversation without human intervention.

\subsection{Rescue}

In the rescue phase, we considered two ways of exploration guidance: \textit {Information Gain-based Rescue} and \textit {Self-resimulation based Rescue}. The former directly provides an explicit exploration guidance, while the latter only blocks wrong exploration paths, allowing dialogue policies to explore diverse paths.

\subsubsection{Information Gain-based Rescue}

The fundamental objective of task-oriented dialogue systems is to help users accomplish their specified tasks, of which acquiring information through dialogue interactions is the first important aspect \cite{geishauser2021does}. Therefore, a crucial goal of the dialogue policy is to maximize access to the user's needs by asking appropriate and targeted questions. On the other side, information gain indicates the amount of information gained about a random variable or signal from observing another random variable \cite{geishauser2021does, padmakumar2021dialog}. Inspired by the usage of information gain in dialog systems \cite{geishauser2021does, padmakumar2021dialog}, we construct an information gain-based rescue approach. As shown in Fig.~\ref{mainb}, this rescue method provides an explicit exploration guidance on the task-oriented database.
To the best of our knowledge, it is the first time to use information gain for this purpose in task-oriented dialog systems.

Specifically, information gain allows a better assessment of the amount of information carried by knowing a certain slot-value. Prioritizing queries for slots with more information gain can help the dialogue policy accomplish user goals faster. Formally, the information gain for each query action \textit{request(slot)} $a$ is as follows:

\vspace{-0.55cm} 
\begin{align}
IG(N,a) &= H(N) - H(N|V_a) 
\\ & = H(N) -  {\textstyle \sum_{v \in V_a}p(N^v)H(N^v|V_a=v)}  \nonumber
\\
H(N) &= -{\textstyle \sum_{n_i \in N} p(n_i) log~p(n_i)}  
\end{align}
\vspace{-0.55cm} 

\noindent where $N$ indicates the entries in the database, $n_i$ indicates the $i$th entry in $N$, $V_a$ indicates the possible values that the user answer after performing action $a$, and $N^v$ indicates the entries in the database filtered by the current information and the newly acquired information $v$ in $V$. 
$H(N)$ and $H(N|V_a)$ denote the information entropy and conditional entropy before and after executing action $a$, respectively. The difference between these two represents the information gain $IG(N,a_i)$ of action $a$.

Since other types of actions (e.g., \textit{inform, common}, etc.) tend to be less useful for filtering entries in DB, the information gains of those types of action are always smaller compared to the \textit{request} type. It leads to the possibility that the dialogue agent is biased toward the \textit{request} action, which is obviously not reasonable. To correct this bias, the type of action needs to be chosen first based on the number of matching entries $n$ and the user action before computing the information gain:

\vspace{-0.35cm} 
\begin{small}
	\begin{align*}
	a^s & = \left\{\begin{matrix}
	request(slot)&if~n>1
	\\
	inform(slot)&if~n  = 1 \&   a^u = request(slot)
	\\
	booking&if~n = 1 \& a^u = request(ticket)
	\\
	common &otherwise,
	\end{matrix}\right.
	\end{align*}
\end{small}
\vspace{-0.35cm}

When $n>1$, it is not enough to target the user's needs based on the current information. The user needs to be asked to provide new information. When $n=1$, the user's needs have been captured, and the dialogue policy can either provide the information that the users want to know or book tickets for them when they want it. In all other cases, the dialogue policy responds based on the user action $a^u$, e.g., user action $a^u$= \textit{bye}, then system action $a^s$=\textit{bye}.

Compared with \cite{geishauser2021does}, the information gain-based rescue considers the variation of the probability distribution over all slots, which covers complete information gain.

\subsubsection{Self-resimulation based Rescue}

We also designed a more free exploration guidance, namely self-resimulation based rescue. As shown in Fig.~\ref{mainc}, rather than directly providing an explicit exploration direction, it shields dialog policies from wrong paths and allows them to explore more diverse directions.

Specifically, in each step, the dialogue policy observes a state $s$ and selects an action $a^s$ using an $\epsilon$-greedy policy, where a random action is selected with $\epsilon$ probability, or otherwise, a greedy action is selected, $a^s=argmax_{a'}Q(s,a';\theta_Q)$, where $Q(s,a';\theta_Q)$ which is the approximated value function that is implemented as a Multi-Layer Perceptron (MLP) with parameter $\theta_Q$. Even if the initial dead-end state has been detected and restored to a safe state, directly using the original Deep Q-Network (DQN) \cite{mnih2015human} without adding additional training always selects the same action due to the same state. Instead, here we fix the probability of selecting the action that leads the conversation to a dead-end to $0$ and then performs a resimulation, i.e., use the dialogue policy to re-select from the action space outside this action.

It is worth noting that the used self-resimulation differs from the dialogue rollout \cite{DBLP:journals/corr/LewisYDPB17}. Dialogue rollout does lookahead for all unexecuted actions in the candidate set, while our approach focuses on the already executed actions. Moreover, it blocks a dead-end path and allows dialogue agents to explore more freely than the information gain-based exploration guidance.

\subsection{Dialogue Data Augment}

Self-reflection avoids humans from making the same mistakes \cite{farthing1992psychology}. Inspired by it, we also added a dialogue data augmentation module to prevent dialogue policies from continuously making the same mistakes. Thus, there are three types of experiences in the experience replay buffer for policy learning:  original experience, rescue experience, and warning experience.

For state $s$, the dialogue agent performs the (original) action $a^o$, observes the updated state $s'$, and receives the reward $r$. If state $s$ is detected as an initial dead-end state, the rescue experience $E^r\leftarrow\{s, a^r, s^r, r^r\}~$ and the warning experience  $E^w\leftarrow\{s,a,s', r+r^w\}$ are stored in the experience replay buffer.
For the rescue experience, it replaces the original action $a^o$ with a rescue action $a^r$, while their corresponding reward and next status update are changed accordingly. For the warning experience, it retains the original action $a^o$ but adds a penalty $r^w$.
Otherwise, only the original experience $E^o\leftarrow\{ s, a^o, s', r\}$ is stored.

\section{Experiments}

Our experiments have four objectives:
1) Verifying the effectiveness and robustness of our DDR (Section~\ref{sec:1} and~\ref{sec:2}); 
2) Investigating the advantages and applicability of two rescue modules (Section~\ref{sec:ana}); 
3) Analyzing the efficiency of the dead-end detection module in our DDR method (Section~\ref{sec:hope});
4) Validating the generality of our DDR method (Section~\ref{sec:gen}).

\begin{table*}[t]
	\centering
	\resizebox{\linewidth}{!}{
		\begin{tabular}{ccccccccccc}
			\hline
			\multirow{2}{*}{Domain}     & \multirow{2}{*}{Agent} & \multicolumn{3}{c}{Epoch = 150}                                                                              & \multicolumn{3}{c }{Epoch = 350}                                                                              & \multicolumn{3}{c }{Epoch = 500}                                                                             \\ \cline{3-11} 
			&                        & \multicolumn{1}{c}{Success $\uparrow$}                & \multicolumn{1}{c }{Reward$\uparrow$}               & Turns$\downarrow$              & \multicolumn{1}{c }{Success$\uparrow$}                & \multicolumn{1}{c}{Reward$\uparrow$}               & Turns $\downarrow$                & \multicolumn{1}{c}{Success$\uparrow$}                & \multicolumn{1}{c }{Reward$\uparrow$}              & Turns $\downarrow$                \\ \hline
			\multirow{5}{*}{Movie}      & DQN                    & \multicolumn{1}{c}{0.1546}          & \multicolumn{1}{c }{-25.51}         & 20.34         & \multicolumn{1}{c }{0.5402}          & \multicolumn{1}{c}{10.15}           & 18.42          & \multicolumn{1}{c }{0.5746}          & \multicolumn{1}{c}{13.30}          & 18.32          \\
			& HER                    & \multicolumn{1}{c}{0.4490}          & \multicolumn{1}{c }{0.38}            & 21.57          & \multicolumn{1}{c}{0.6388}          & \multicolumn{1}{c }{20.05}          & 16.38          & \multicolumn{1}{c }{0.7104}          & \multicolumn{1}{c}{27.57}         & 14.24          \\ 
			& LCPO              & 0.4108 & -1.22 & 21.33 & 0.6024 & 19.85 & 16.12 & 0.7258 & 29.74 & 14.35        \\ \cdashline{2-11} 
			& DDR-IG                & \multicolumn{1}{c}{0.4731}          & \multicolumn{1}{c }{3.02}            & \textbf{18.04} & \multicolumn{1}{c  }{{\color[HTML]{3531FF}\textbf{0.7567}}} & \multicolumn{1}{c }{\textbf{31.92}}  & \textbf{12.30} & \multicolumn{1}{c }{{\color[HTML]{3531FF}\textbf{0.8021}}} & \multicolumn{1}{c }{\textbf{36.37}} & \textbf{11.34} \\
			& DDR-SE                & \multicolumn{1}{c}{{\color[HTML]{3531FF} \textbf{0.5329}}} & \multicolumn{1}{c }{\textbf{8.53}}   & 20.36          & \multicolumn{1}{c }{0.7115}          & \multicolumn{1}{c}{27.27}          & 15.02          & \multicolumn{1}{c }{0.7585}          & \multicolumn{1}{c}{32.17}          & 13.70          \\ 
			\hline
			
			\multirow{5}{*}{Restaurant} & DQN                    & \multicolumn{1}{c}{0.0530}          & \multicolumn{1}{c}{-39.30}          & 30.15          & \multicolumn{1}{c}{0.1971}          & \multicolumn{1}{c}{-24.22}         & 25.92          & \multicolumn{1}{c}{0.4110}          & \multicolumn{1}{c}{-3.52}         & 23.02          \\ 
			& HER                    & \multicolumn{1}{c}{0.1022}          & \multicolumn{1}{c}{-33.70}          & 27.79          & \multicolumn{1}{c}{0.4840}          & \multicolumn{1}{c}{4.01}           & 21.10           & \multicolumn{1}{c}{0.6664}          & \multicolumn{1}{c}{22.10}         & 17.75          \\
			& LCPO        &0.0903 & -35.65 & 27.64 & 0.4831 & 3.74 & 21.30 & 0.6949 & 27.69 & 16.94    \\ \cdashline{2-11} 
			& DDR-IG                & \multicolumn{1}{c}{0.1038}          & \multicolumn{1}{c}{-33.26}          & \textbf{26.22} & \multicolumn{1}{c}{{\color[HTML]{3531FF}\textbf{0.6230}}} & \multicolumn{1}{c}{\textbf{17.32}} & \textbf{17.79} & \multicolumn{1}{c}{0.7582}          & \multicolumn{1}{c}{30.74}          & \textbf{15.44} \\
			& DDR-SE                & \multicolumn{1}{c}{{\color[HTML]{3531FF}\textbf{0.1064}}} & \multicolumn{1}{c}{\textbf{-32.79}} & 26.73         & \multicolumn{1}{c}{0.4970}          & \multicolumn{1}{c}{4.82}           & 21.83          & \multicolumn{1}{c}{{\color[HTML]{3531FF}\textbf{0.7716}}} & \multicolumn{1}{c}{\textbf{32.25}} & 16.38        \\ 
			\hline

			\multirow{5}{*}{Taxi}       & DQN               & \multicolumn{1}{c}{0.0548}          & \multicolumn{1}{c}{-37.5}           & 26.85          & \multicolumn{1}{c}{0.2707}          & \multicolumn{1}{c}{-16.05}          & 22.82           & \multicolumn{1}{c}{0.6300}          & \multicolumn{1}{c}{18.31}         & 18.78          \\ 
			& HER               & \multicolumn{1}{c}{0.0142}          & \multicolumn{1}{c}{-41.72}          & 28.00          & \multicolumn{1}{c}{0.0811}          & \multicolumn{1}{c}{-35.18}          & 26.95          & \multicolumn{1}{c}{0.1259}          & \multicolumn{1}{c}{-30.72}        & 26.10          \\
			& LCPO           & 0.0674 & 36.65 & 27.79 & 0.3565 & 5.38 & 21.90 & 0.6404 & 19.69 & 19.01     \\ \cdashline{2-11} 
			& DDR-IG             & \multicolumn{1}{c}{{\color[HTML]{3531FF}\textbf{0.0984}}}          & \multicolumn{1}{c}{\textbf{-33.96}}          & \textbf{25.14} & \multicolumn{1}{c}{{\color[HTML]{3531FF}\textbf{0.5299}}} & \multicolumn{1}{c}{\textbf{9.55}}  & \textbf{16.83} & \multicolumn{1}{c}{{\color[HTML]{3531FF}\textbf{0.7821}}} & \multicolumn{1}{c}{\textbf{32.90}} & \textbf{14.85} \\ 
			& DDR-SE            & \multicolumn{1}{c}{0.0790}          & \multicolumn{1}{c}{-35.57}          & 27.35          & \multicolumn{1}{c}{0.3974}          & \multicolumn{1}{c}{-3.21}          & 19.95          & \multicolumn{1}{c}{0.6654}          & \multicolumn{1}{c}{21.64}         & 18.49          \\ 
			\hline
		\end{tabular}
	}
	\caption{Results of different agents on three datasets in the normal environment. The highest performance in each column is highlighted. Each result is averaged over 10 turns out of 1000 dialogues using different random seeds. The difference between the results of all agent pairs evaluated at the same epoch is statistically significant ($p < 0.05$).} 
	\label{tab:main}
\end{table*}

\subsection{Settings}

Experiments have been conducted to evaluate the objectives using a dialogue simulation environment with an error model \cite{li2016user}. 
We conducted the experiments on the Microsoft dialogue challenge, which provided three standard datasets and a unified experimental environment to collaborate and benchmark for the dialogue research community \cite{li2018microsoft}. We also performed further validation on the Multiwoz dataset \cite{budzianowski2018multiwoz}. The detail of the main metrics and datasets are in appendices ~\ref{ap:metrics} and~\ref{ap:datasets}.

For a fair comparison, all dialogue agents are trained by DQN algorithm (except the generalization experiments in Sec.~\ref{sec:gen}) with the same sets of hyperparameters. Each policy network has one hidden layer with 80 hidden nodes. The learning rate is $0.001$, and the batch size is $16$. The optimizer is Adam \cite{kingma2014adam}. $\epsilon$-greedy is always applied for exploration with $0.1$ and decayed to $0.01$ during training.  The buffer size of $D$ is $10000$. The number of warm start epochs is $120$. 
It is not reasonable to make a re-decision whenever an erroneous action is detected, thus we set 
the maximum number of recoveries to $3$ for our method. Each result is averaged over ten turns on 1000 dialogues with random seeds. 
For the reward function, a big bonus of $2L$ for success and a penalty of $-L$ for failure. Otherwise, a fixed penalty of $-1$ is provided for each turn to encourage short interactions. All agents are limited to thirty turns, $L = 30$, and their discount factor is $0.95$.

\subsection{Baselines}

We compare with different dialogue agents that make RL effective and efficient by similarly manipulating dialogue trajectories as a baseline \footnote{It is worth noting that we do not consider methods that use additional human intervention since an important advantage of our approach is that it does not require any human intervention.}:

\textbf{DQN} agent learned with one standard Deep Q-network \cite{mnih2015human}.

\textbf{HER} agent is learned from failure by generating artificial successful dialogue instances for DDA  \cite{lu2019goal}.

\textbf{LCPO} agent is learned from clipped dialogue trajectories \footnote{Since \citet{wu2021clipping} demonstrated the superiority of the LCPO compared to TAM \cite{wu2019tam}, we only use the better-performing LCPO as the baseline model.}\cite{wu2021clipping}.

\textbf{Proposed DDR}

\textbf{DDR-IG} agent is learned with a DQN that incorporates proposed dead-end detection, information gain-based rescue, and dialogue data augment. 

\textbf{DDR-SE} agent is similar to DDR-IG except that the rescue module is based on self-resimulation.

\begin{table}[t]
	\centering
	\resizebox{.8\linewidth}{!}{
		\begin{tabular}{cccc}
			\hline
			Agents      & Success    $\uparrow$             & Reward      $\uparrow$           & Turns $\downarrow$              \\ \hline
			DQN& 0.0726          & -48.83           & 37.08         \\ 
			HER & 0.0862          & -47.04           & 36.76          \\
			LCPO& 0.1008          & -40.71          & 34.53          \\ \hdashline
			DDR-SE& 0.1793          & -33.85          & 32.72          \\ 
			DDR-IG& {\color[HTML]{3531FF}\textbf{0.2010}} & \textbf{-31.01} & \textbf{32.27} \\ \hline
		\end{tabular}
	}
	\caption{Results of different agents on the Multiwoz dataset.} 
	\label{tab:multi}
\end{table}

\begin{table*}[t]
	\centering
	\resizebox{\linewidth}{!}{
		\begin{tabular}{ccccccccccc}
			\hline
			\multirow{2}{*}{Domain}     & \multirow{2}{*}{Agent} & \multicolumn{3}{c}{Epoch = 150}                                                                              & \multicolumn{3}{c}{Epoch = 350}                                                                               & \multicolumn{3}{c}{Epoch = 500}                                                                              \\ \cline{3-11} 
			&                        & \multicolumn{1}{c}{Success $\uparrow$}                & \multicolumn{1}{c}{Reward $\uparrow$}               & Turns $\downarrow$              & \multicolumn{1}{c}{Success $\uparrow$}                & \multicolumn{1}{c}{Reward $\uparrow$}                & Turns  $\downarrow$             & \multicolumn{1}{c}{Success $\uparrow$}                & \multicolumn{1}{c}{Reward $\uparrow$}               & Turns$\downarrow$               \\ \hline
			\multirow{5}{*}{Movie}      & DQN                    & \multicolumn{1}{c}{0.2917}          & \multicolumn{1}{c}{-15.04}          & 24.10          & \multicolumn{1}{c}{0.4405}          & \multicolumn{1}{c}{0.75}             & 19.30          & \multicolumn{1}{c}{0.5086}          & \multicolumn{1}{c}{7.70}            & 17.65          \\ 
			& HER                    & \multicolumn{1}{c}{0.0845}          & \multicolumn{1}{c}{-33.59}          & 23.90         & \multicolumn{1}{c}{0.4948}          & \multicolumn{1}{c}{5.79}             & 18.99          & \multicolumn{1}{c}{0.5584}          & \multicolumn{1}{c}{12.00}          & 18.02          \\ 
			& LCPO            &0.3123 & -11.20 & 23.12 & 0.5033 & 10.62 & 18.80 & 0.5902 & 15.37 & 16.28      \\  \cdashline{2-11}
			& DDR-IG                & \multicolumn{1}{c}{0.3428}          & \multicolumn{1}{c}{-9.59}           & \textbf{20.44} & \multicolumn{1}{c}{0.6247}          & \multicolumn{1}{c}{18.45}            & \textbf{14.33} & \multicolumn{1}{c}{0.6217}          & \multicolumn{1}{c}{18.06}           & \textbf{13.96} \\
			& DDR-SE                & \multicolumn{1}{c}{{\color[HTML]{3531FF}\textbf{0.4297}}} & \multicolumn{1}{c}{\textbf{-1.22}} & 21.29          & \multicolumn{1}{c}{{\color[HTML]{3531FF}\textbf{0.6450}}} & \multicolumn{1}{c}{\textbf{21.03}}   & 15.54          & \multicolumn{1}{c}{{\color[HTML]{3531FF}\textbf{0.6974}}} & \multicolumn{1}{c}{\textbf{26.28}}  & 14.47          \\ 
			\hline

			\multirow{5}{*}{Restaurant} & DQN                    & \multicolumn{1}{c}{0.0486}          & \multicolumn{1}{c}{-38.25}          & 27.25          & \multicolumn{1}{c}{0.2927}          & \multicolumn{1}{c}{-14.69}           & 24.06          & \multicolumn{1}{c}{0.4226}          & \multicolumn{1}{c}{-1.99}          & 22.05          \\

			& HER                    & \multicolumn{1}{c}{0.0507}          & \multicolumn{1}{c}{-38.66}          & 28.45          & \multicolumn{1}{c}{0.1785}          & \multicolumn{1}{c}{-26.45}           & 27.02          & \multicolumn{1}{c}{0.3501}          & \multicolumn{1}{c}{-9.32}          & 23.66          \\ 
			& LCPO            & 0.0655 & -35.54 & 27.01 & 0.2954 & -15.25 & 23.11 & 0.4292 & -0.41 & 20.84        \\ \cdashline{2-11} 
			& DDR-IG                & \multicolumn{1}{c}{0.0757}          & \multicolumn{1}{c}{-36.23}          & \textbf{25.45}          & \multicolumn{1}{c}{0.3999}          & \multicolumn{1}{c}{-4.85}           & \textbf{20.19} & \multicolumn{1}{c}{0.4354}          & \multicolumn{1}{c}{-1.40}          & \textbf{18.29} \\ 
			& DDR-SE                & \multicolumn{1}{c}{{\color[HTML]{3531FF} \textbf{0.1706}}}          & \multicolumn{1}{c}{\textbf{-27.49}}          & 26.44          & \multicolumn{1}{c}{{\color[HTML]{3531FF}\textbf{0.5020}}} & \multicolumn{1}{c}{\textbf{5.30}}    & 20.65          & \multicolumn{1}{c}{{\color[HTML]{3531FF}\textbf{0.5880}}} & \multicolumn{1}{c}{\textbf{13.96}} & 18.66          \\ 
			\hline

			\multirow{5}{*}{Taxi}       & DQN               & \multicolumn{1}{c}{0.0848}          & \multicolumn{1}{c}{-35.25}          & 27.76          & \multicolumn{1}{c}{0.2142}          & \multicolumn{1}{c}{-21.85}          & 24.24       & \multicolumn{1}{c}{0.4682}          & \multicolumn{1}{c}{2.64}           & 21.00          \\
			& HER               & \multicolumn{1}{c}{0.0278}          & \multicolumn{1}{c}{-41.34}          & 29.68         & \multicolumn{1}{c}{0.1403}          & \multicolumn{1}{c}{-29.98}          & 27.21          & \multicolumn{1}{c}{0.3164}          & \multicolumn{1}{c}{-12.32}         & 23.59          \\
			& LCPO      &  0.0645 & -40.29 & 29.81 & 0.1702 & -26.51 & 26.67 & 0.4019 & -7.36 & 22.10  \\  \cdashline{2-11}
			& DDR-IG             & \multicolumn{1}{c}{0.0276}          & \multicolumn{1}{c}{-41.19}          & \textbf{24.15} & \multicolumn{1}{c}{0.0934}          & \multicolumn{1}{c}{-34.09}           & 22.04          & \multicolumn{1}{c}{0.2807}          & \multicolumn{1}{c}{-15.97}         & 21.30          \\
			& DDR-SE           & \multicolumn{1}{c}{{\color[HTML]{3531FF}\textbf{0.0860}}} & \multicolumn{1}{c}{\textbf{-34.21}} & 25.91         & \multicolumn{1}{c}{{\color[HTML]{3531FF}\textbf{0.3163}}} & \multicolumn{1}{c}{\textbf{-11.35}} & \textbf{21.64}          & \multicolumn{1}{c}{{\color[HTML]{3531FF}\textbf{0.6017}}} & \multicolumn{1}{c}{\textbf{15.43}} & \textbf{19.45} \\ 
			\hline
		\end{tabular}
	}
	\caption{Results of different agents on three datasets in noisy environment. The highest performance in each column is highlighted. Each result is averaged over 10 turns out of 1000 dialogues using different random seeds. The difference between results of all agent pairs evaluated at the same epoch is statistically significant ($p < 0.05$).} 
	\label{tab:noisy}
\end{table*}

\subsection{Effectiveness Evaluation}
\label{sec:1}
To verify the effectiveness of our method, we conduct experiments on three single-domain datasets in a normal environment (without noise). Table~\ref{tab:main} reports the main result of different agents. The results show that our two methods outperform other baselines with a statistically significant margin. Compared to DQN, HER shows significant improvement in the movie and restaurant domains, while it performs poorly in the taxi domain. This is due to the fact that the effectiveness of HER is highly affected by the quality of successful artificial dialogues, and in the relatively complex taxi domain, low-quality artificial successful dialogues for DDA hurt its performance. LCPO makes the dialogue samples more effective by clipping lengthy dialogue trajectories. However, avoiding invalid exploration produced by dead-ends is still tricky. In contrast, our two DDR methods make exploration more effective and diverse by detecting dead-ends and rescuing and guiding them promptly.

We further validate the performance of our two DDR methods on a multi-domain dataset. Each agent was trained on this dataset for 1000 epochs and updated every $200$ experience collected. Likewise, each result was averaged over 10 runs using different random seeds. Table~\ref{tab:multi} shows the performance of the different agents without noise. Consistent with the single domain results, better performance is achieved in two DDR methods.

\begin{table}[t]
	\centering
	\resizebox{1.\linewidth}{!}{
		\begin{tabular}{cccc}
			\hline
			Agent     & Movie                  & Restaurant             & Taxi                   \\ \hline
			DQN       & 64.29\%$\pm$11.06\%        & 82.58\%$\pm$5.06\%         & 88.36\%$\pm$5.58\%         \\
			HER       & 37.35\%$\pm$9.88\%         & 56.19\%$\pm$10.33\%        & 71.28\%$\pm$9.18\%         \\ 
			LCPO & 36.22\%$\pm$8.09\%        & 44.50\%$\pm$4.26\%         & 50.88\%$\pm$5.57\%         \\ \hdashline
			DDR-SE   & 23.25\%$\pm$6.34\%         & 30.04\%$\pm$3.31\%         & 29.22\%$\pm$2.09\%         \\ 
			DDR-IG   & {\color[HTML]{3531FF}\textbf{3.32\%$\pm$0.75\%}} &{\color[HTML]{3531FF} \textbf{7.40\%$\pm$1.82\%} }& {\color[HTML]{3531FF}\textbf{8.21\%$\pm$1.15\%}} \\ \hline
		\end{tabular}
	}
	\caption{Statistics on the percentage of failed conversations with dead-ends for different agents in three domains. $\pm$ is the mean standard deviation of the average of 10 runs with different seeds.} 
	\label{tab:hope}
\end{table}

\subsection{Robustness Evaluation}
\label{sec:2}

The noise in the real-life dialogue environment contains the accumulation error of upstream modules and ambiguous user utterances, exacerbating the problem of dead-ends in the dialogue \cite{ZhaoWYZHW20}. A dialogue system should be able to carry on a conversation without the luxury of accurate upstream modules or precise user utterances \cite{PaekH00}. Therefore, to verify the robustness of our approach, we constructed experiments in noisy environments with and 10\% slot error rate. The results are shown in Table~\ref{tab:noisy}, and our method also yields better performance except for DDR-IG on the relatively difficult taxi domain (We further analyze the reasons for it in section~\ref{sec:ana}).

\begin{figure*}[t]
	\centering
	\subcaptionbox{User goal 34}{\includegraphics[width=0.64\columnwidth]{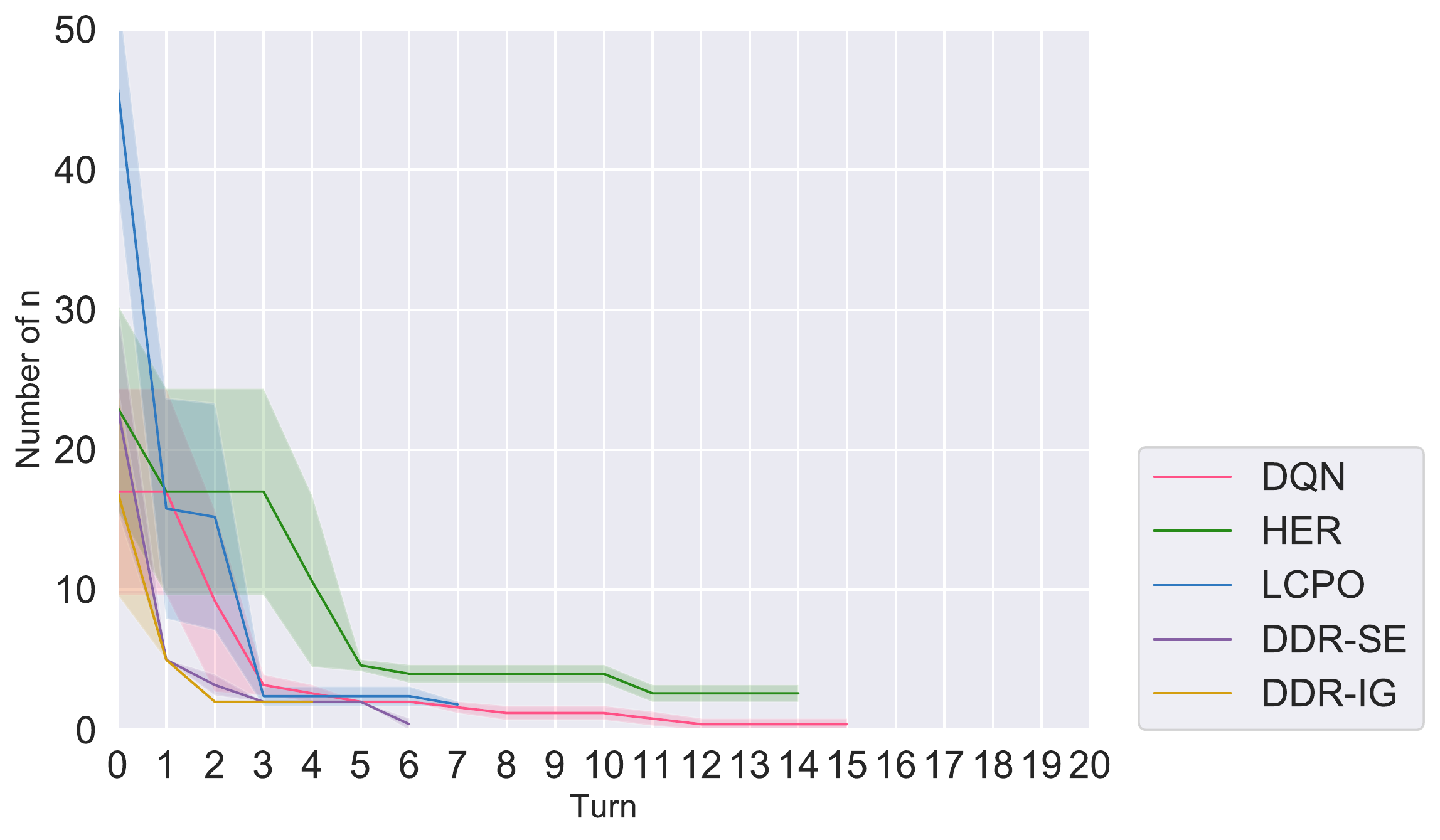}}
	\subcaptionbox{User goal 97}{\includegraphics[width=0.64\columnwidth]{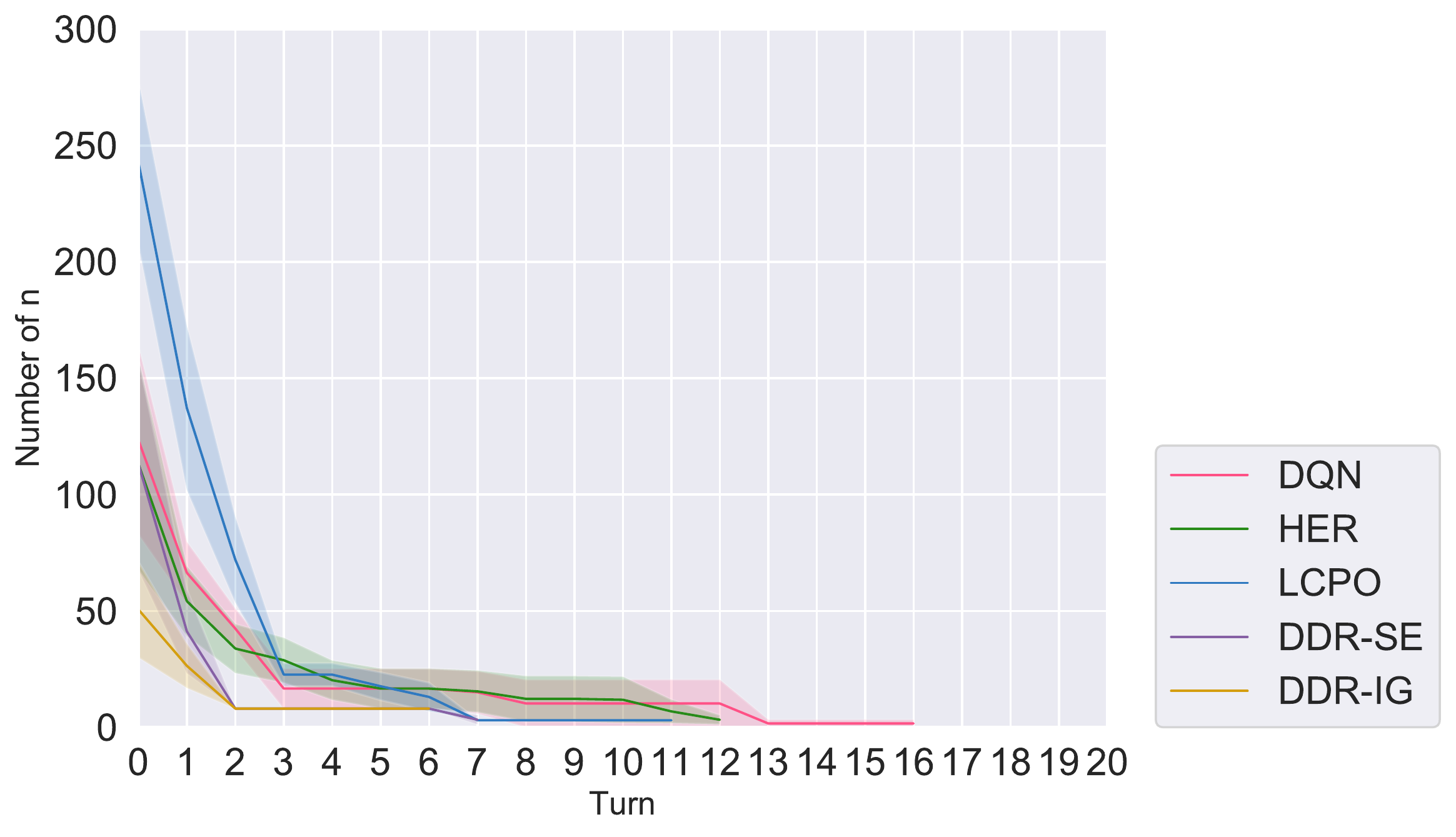}}
	\subcaptionbox{User goal 121}{\includegraphics[width=0.64\columnwidth]{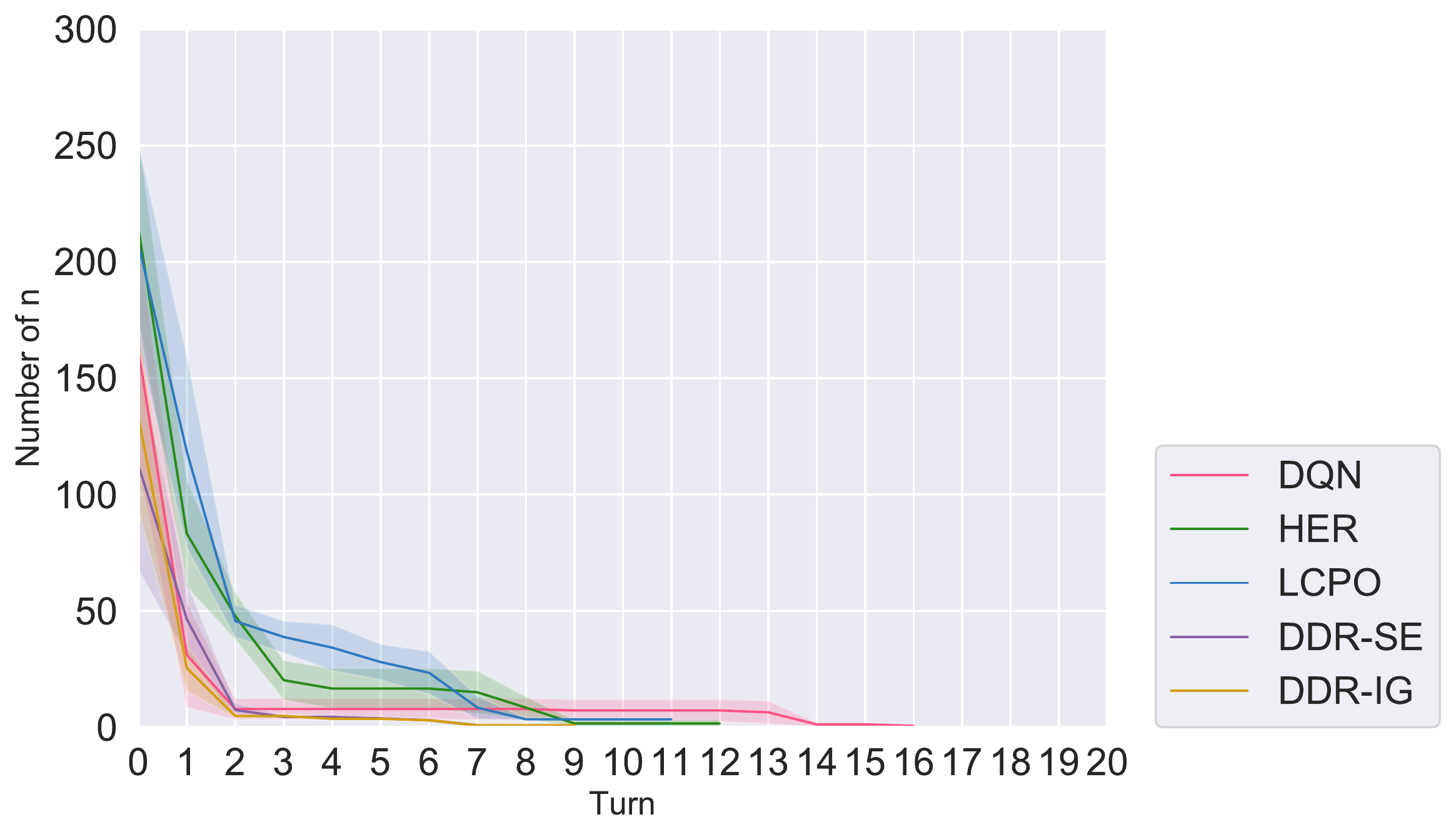}}
	\caption{Visualization of the variation of $n$ with turns of interaction for different user goals.  All results are the average values from 10 runs with different random seeds.}
	\label{fig:n}
\end{figure*}

\begin{figure*}[t]
	\centering
	\subcaptionbox{Distribution DQN}{\includegraphics[width=0.64\columnwidth]{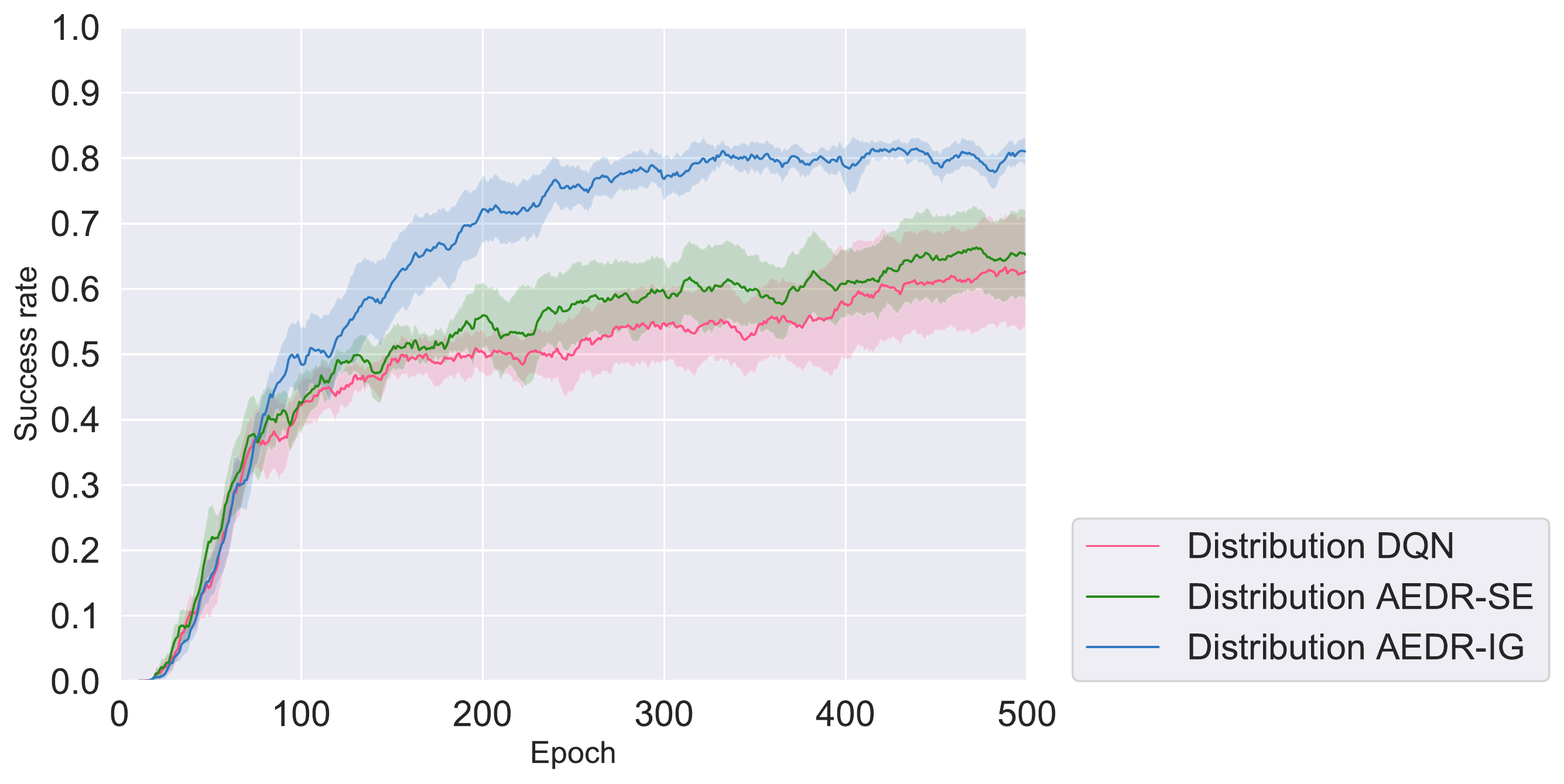}}
	\subcaptionbox{Double DQN}{\includegraphics[width=0.64\columnwidth]{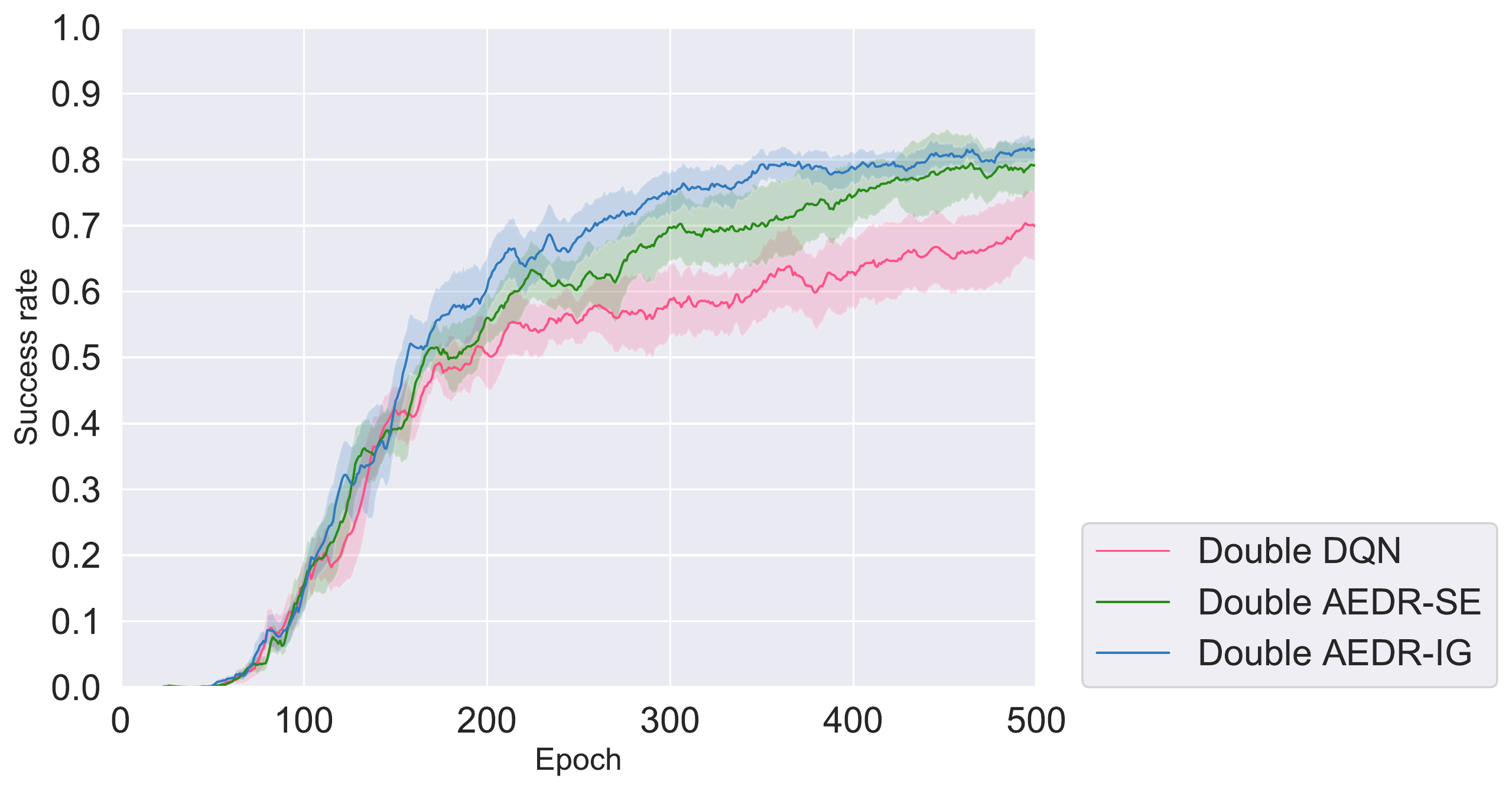}}
	\subcaptionbox{Dueling DQN}{\includegraphics[width=0.64\columnwidth]{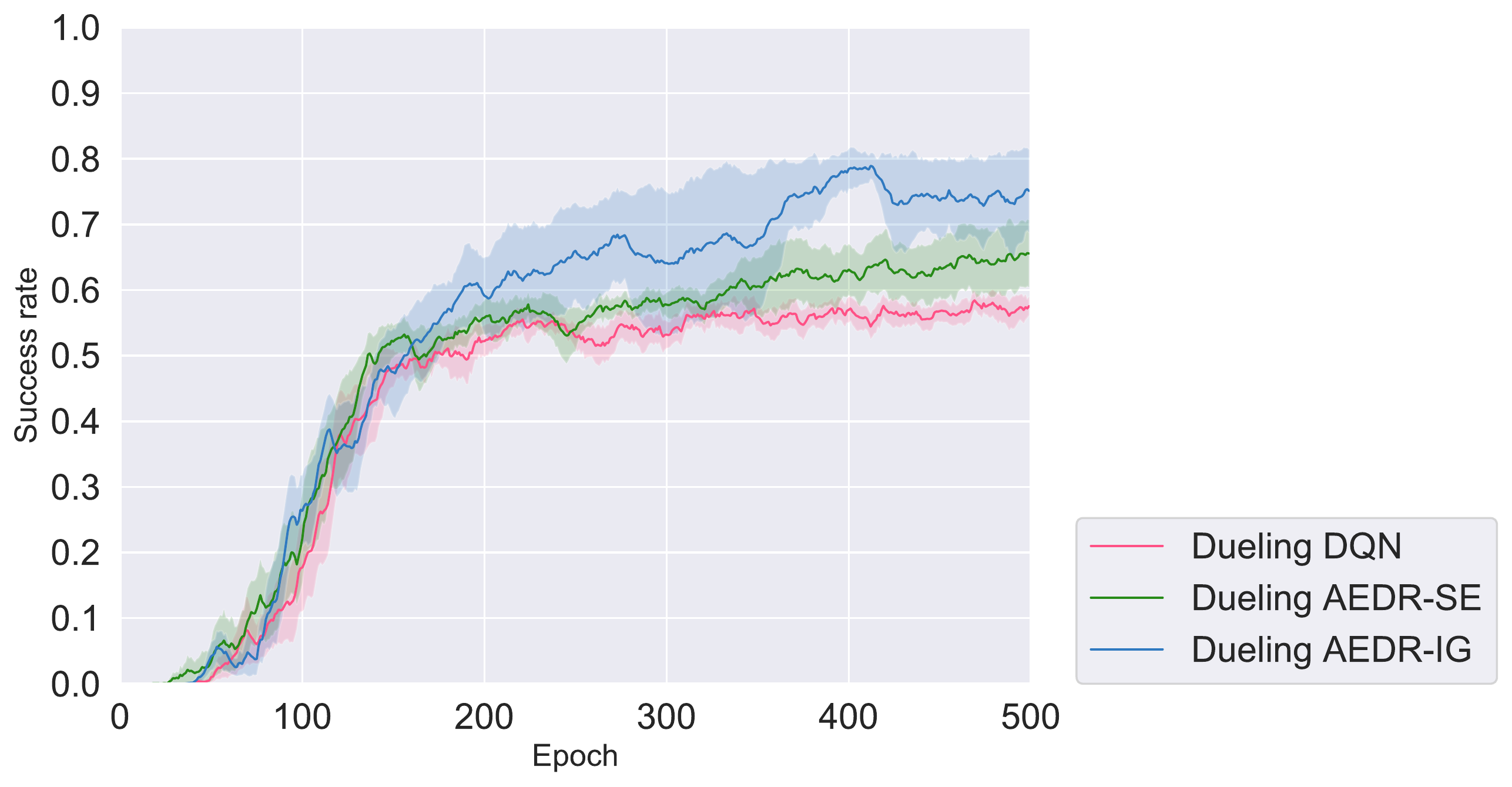}}
	\caption{Incorporating different variants of DQN with our two methods on movie domain.}
	\label{fig:m}
\end{figure*}

\begin{figure*}[!t]
	\centering
	\subcaptionbox{Distribution DQN}{\includegraphics[width=0.64\columnwidth]{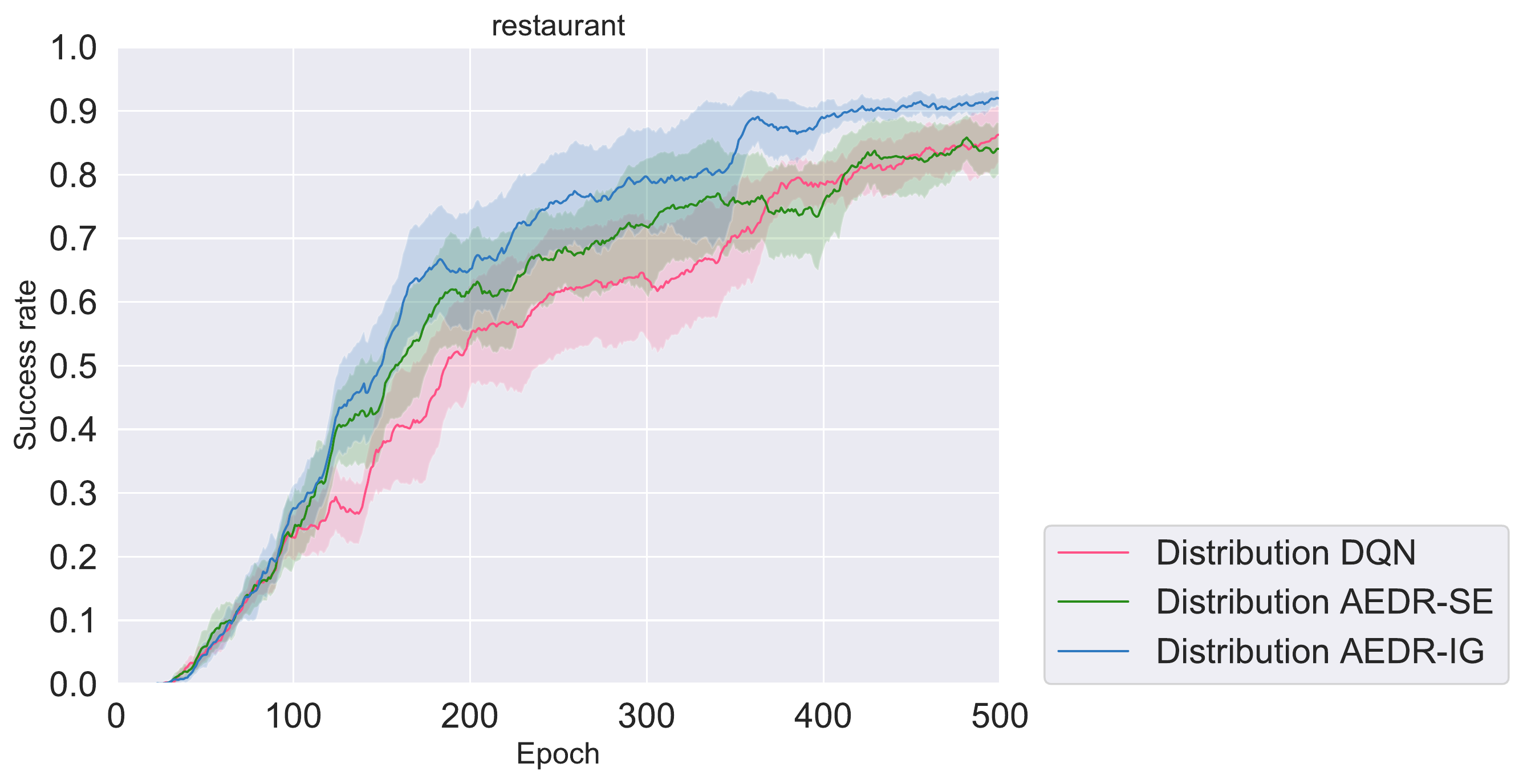}}
	\subcaptionbox{Double DQN}{\includegraphics[width=0.64\columnwidth]{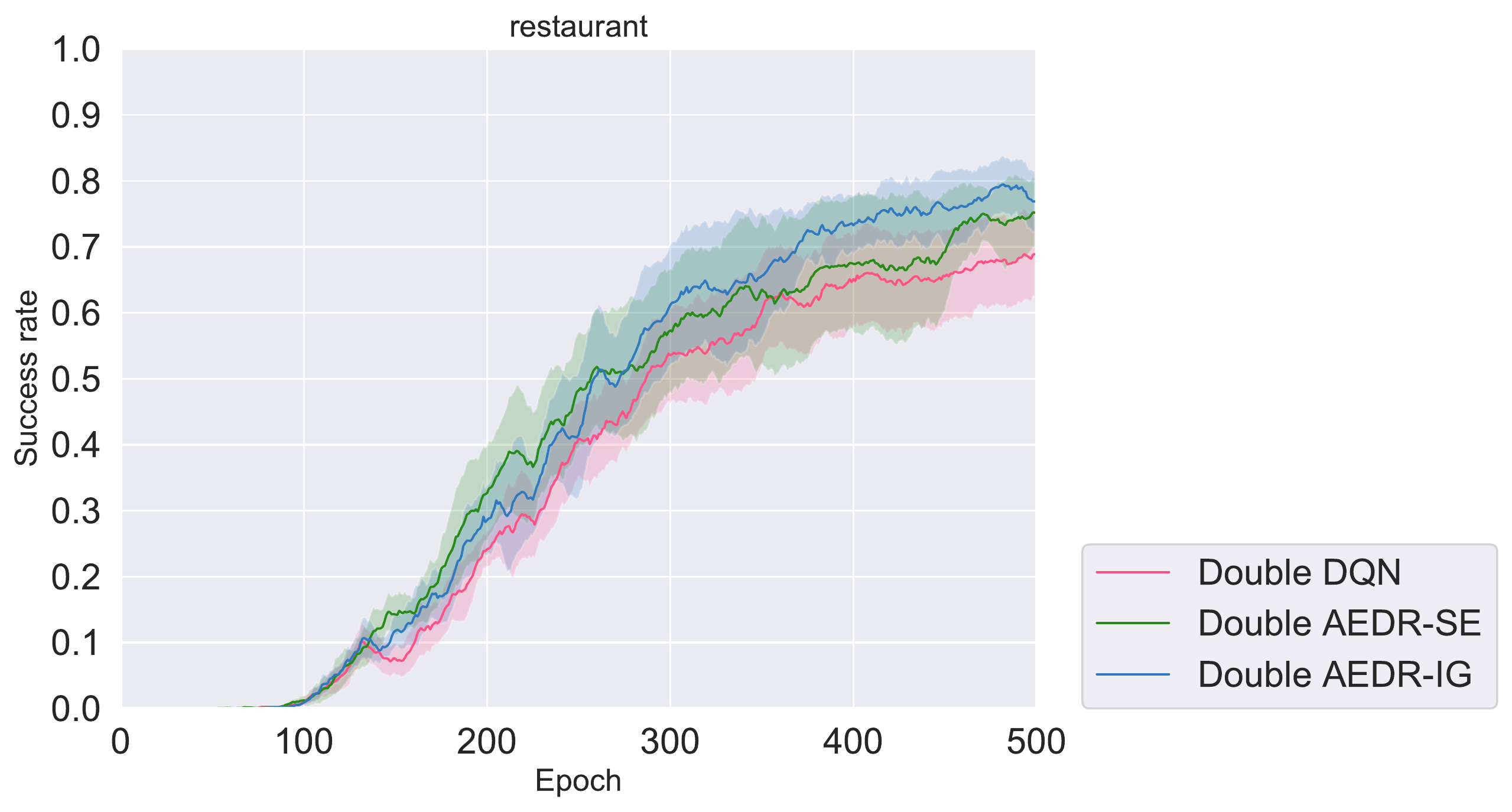}}
	\subcaptionbox{Dueling DQN}{\includegraphics[width=0.64\columnwidth]{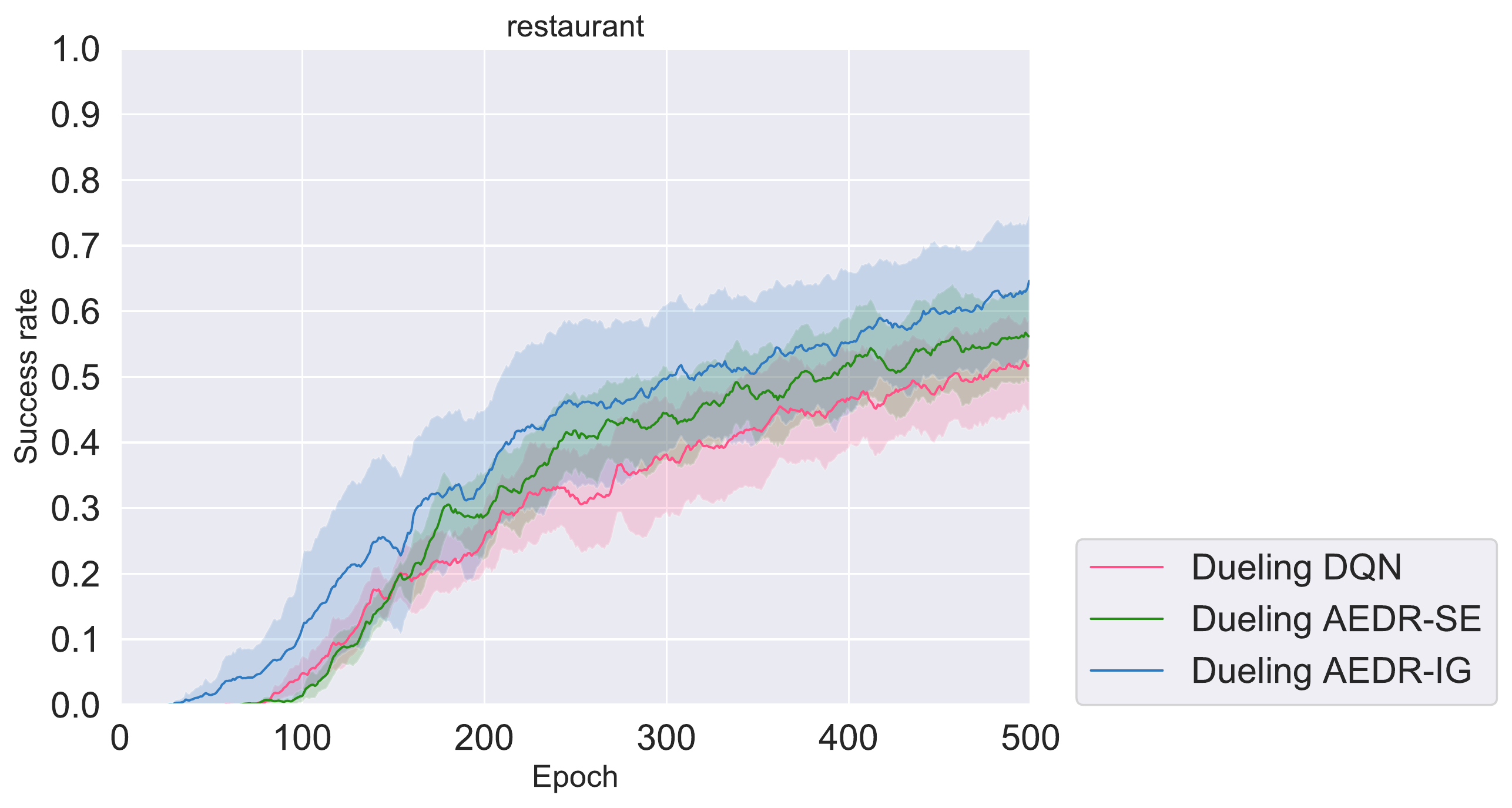}}
	\caption{Incorporating different variants of DQN with our two methods on resturant domain.}
	\label{fig:r}
\end{figure*}

\begin{figure*}[!t]
	\centering
	\subcaptionbox{Distribution DQN}{\includegraphics[width=0.64\columnwidth]{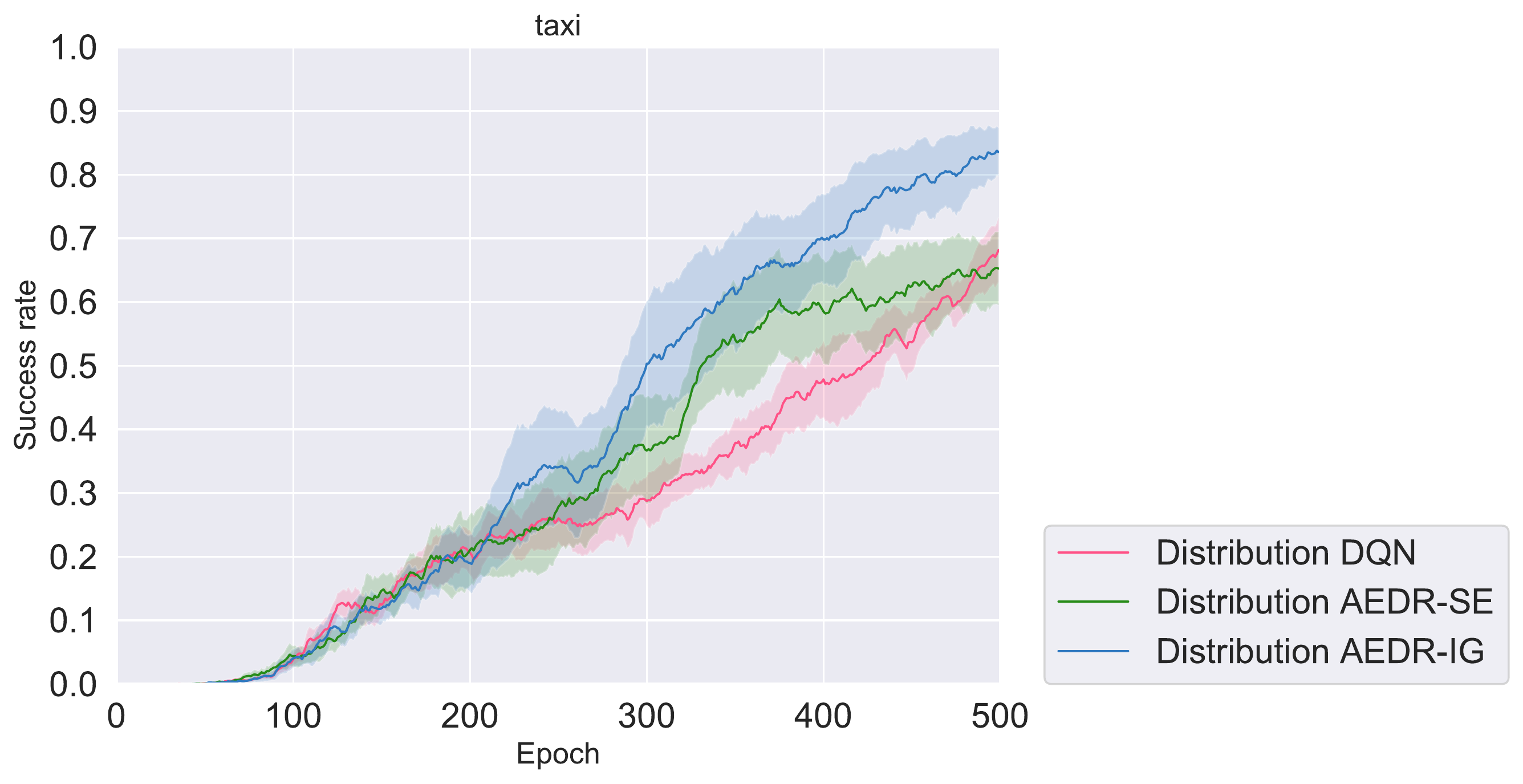}}
	\subcaptionbox{Double DQN}{\includegraphics[width=0.64\columnwidth]{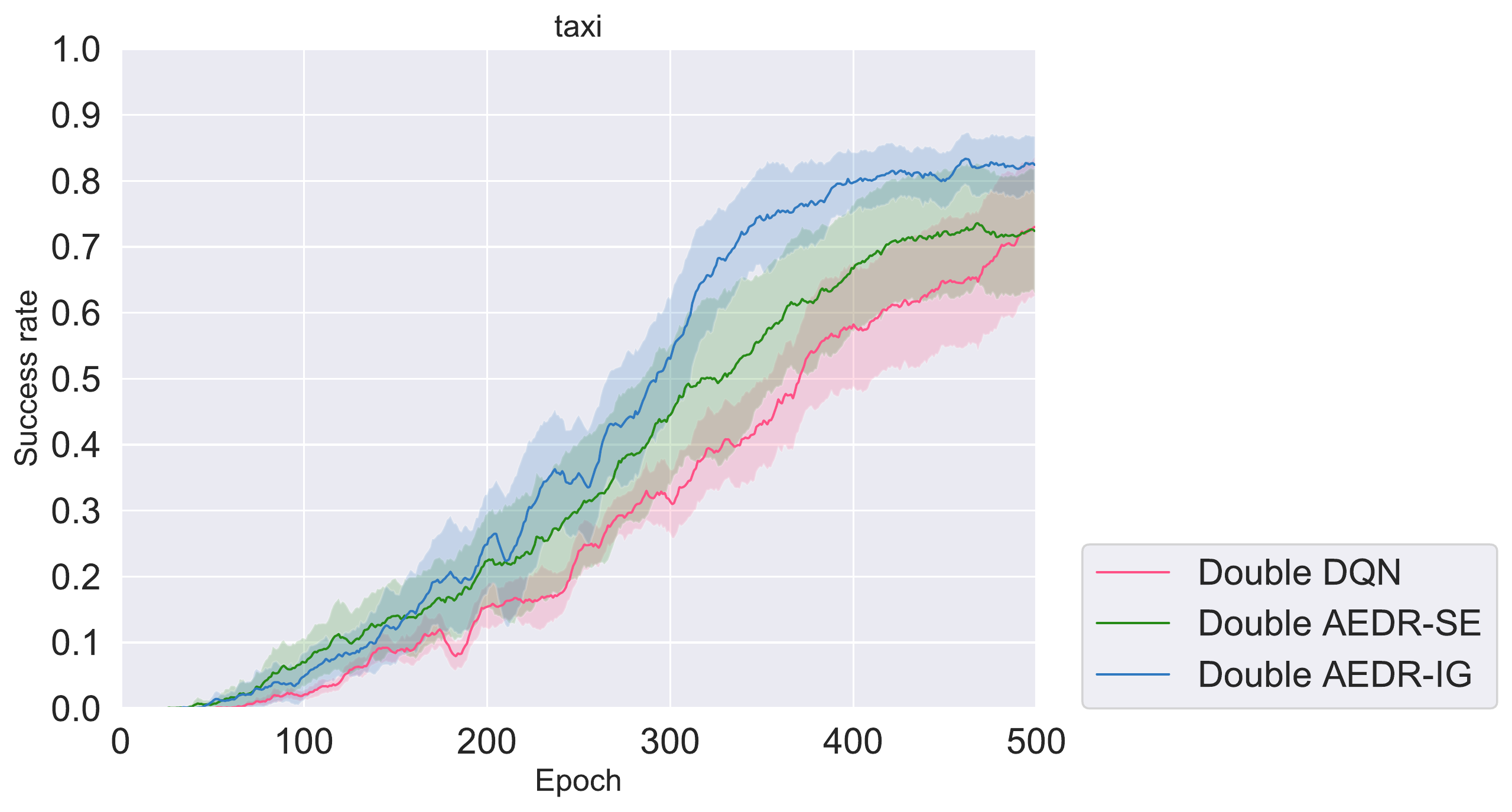}}
	\subcaptionbox{Dueling DQN}{\includegraphics[width=0.64\columnwidth]{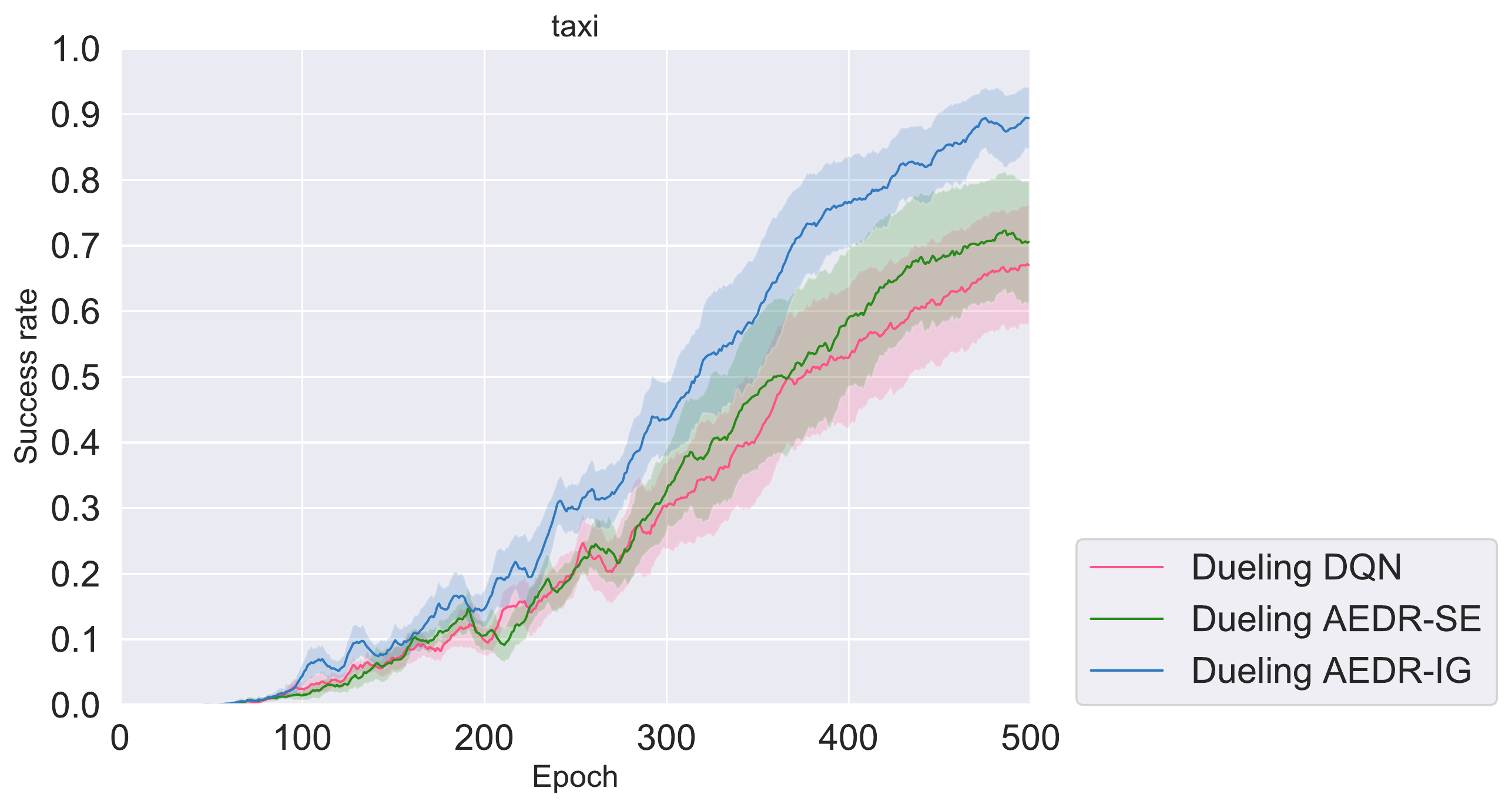}}
	\caption{Incorporating different variants of DQN with our two methods on taxi domain.}
	\label{fig:t}
\end{figure*}

\subsection{Effectiveness of Different DDR Methods}
\label{sec:ana}

We observed different performances of our two methods in noisy and normal (noise-free) environments. DDR-IG performs best in the noise-free environment, while DDR-SE performs best in the noisy environment. We believe that using information gain to select actions avoids suboptimal actions, thus making DDR-IG more efficient than DDR-SE in a noise-free environment. At the same time, the usability of DDR-IG is limited by the inability to accurately estimate information gains due to noise, while DDR-SE is able to reduce the effect of noise through exploration more freely.

\subsection{Effectiveness of DDR for Dead-ends}
\label{sec:hope}


We counted the percentage of dead-ends conversations among the failed conversations for different dialogue agents. Higher percentages mean that more low-quality conversations containing dead-end are generated. As shown in Tab.~\ref{tab:hope}, two DDR methods significantly alleviate this problem. It shows that our two DDR methods accurately identify dead-end with high probability and rescue them instead of continuing meaningless conversations.

Moreover, we randomly selected three user goals in the movie domain to visualize the variation of $n$ with turns for different dialogue agents, as shown in Fig.~\ref{fig:n}. When n quickly shrinks to a non-zero value and disappears, the dialogue agent quickly targets the matching entries by efficient interaction with the user and achieves the user goal. It can be seen that our two approaches always capture user needs faster through effective conversations and target matching entries to accomplish user goals. At the same time, the other methods continue the meaningless conversation even if the conversation reaches a dead-end.

\subsection{Generality Evaluation}
\label{sec:gen}

To verify the generalizability of our method, we combine our two methods with different value-based RL models in three domains. Fig.~\ref{fig:m}-\ref{fig:t} show their learning curves in three domains. We can observe that combining both of our methods improves their learning efficiency to different degrees.
We can conclude that our method generalizes to most RL-based dialogue policy algorithms.

\section{Conclusion}

This paper finds and defines one of the critical reasons for the invalid exploration of dialogue policies, - dead-ends -, and proposes a novel dead-end resurrection (DDR) algorithm. The algorithm effectively rescues conversations from dead-ends and provides a rescue action to guide and correct the exploration direction. Dead-end detection in DDR introduces an efficient criterion for detecting dead-ends, and its effectiveness has been experimentally verified. Once a certain dead-end is detected, rescue in DDR provides rescue exploration guidance for avoiding invalid exploration and thus generates more high-quality and diverse conversations. In addition, the experience containing dead-ends is added for DDA to prevent the same mistakes from occurring in the dialogue policy. The extensive results demonstrate the superiority and generality of our approach.
In the future, we plan to explore more different rescue algorithms and combine them with the dead-end detection module to achieve various desired properties for the exploration.

\section*{Limitations}

Although our DDR algorithm is efficient and very general to be used with different RL algorithms on task-oriented dialogue systems that work on the database, it is difficult to use for other types of dialogue systems where the database is not accessible. Even though this paper focuses on task-oriented dialogue systems that work on the database, it is undeniable that other types of dialogue systems with hard-to-access databases may also have dead-end problems. In the future, we will also work on these database-inaccessible dialogue systems to further address this issue.

\bibliography{anthology,custom}
\bibliographystyle{acl_natbib}

\clearpage
\appendix
\appendix

\section{Main Metrics}
\label{ap:metrics}
To verify the effectiveness of the proposed method, three metrics are used to evaluate the performance of our DDR: Success Rate (SR), Average Reward (AE), and Average Turns (AT). 

\begin{itemize}
	\item SR: the fraction of dialogs that finish successfully in multiple experiments. A dialogue is considered successful only if all user requests have been informed and the booked entities satisfy the constraint. A higher SR indicates a better overall dialogue policy.
	\begin{eqnarray}
	SR & = & \frac{N_{succ}}{N_{total}} 
	\end{eqnarray}
	\noindent where $N_{total}$ indicates the total number of dialogues and $N_{succ}$ indicates the number of successful dialogues.
	\item  AE: the average reward received during the dialogue. A good policy should have a high everage AE.
	\begin{eqnarray}
	AE = \frac {\textstyle \sum_{i=1}^{N_{total}}R_i} {N_{total}}
	\end{eqnarray}
	\noindent where $R_i$ denotes the cumulative return of the $i-th$ dialogue.
	\item AT: the average length of the dialogue. A fewer AT indicate fewer turns required to complete the task and more efficient of the dialogue policy.
	\begin{eqnarray}
	AE = \frac {\textstyle \sum_{i=1}^{N_{total}}L_i} {N_{total}}
	\end{eqnarray}
	\noindent where $L_i$ denotes the length of the $i-th$ dialogue.
	
\end{itemize}

There is a strong correlation among the three metrics: generally speaking, a good policy should have a high success rate, high average reward and low average turns. Among them, SR is the major evaluation metric \cite{li2018microsoft}.

\section{Datasets}
\label{ap:datasets}

Table~\ref{tab:dataset} demonstrates the full statistics of comparisons between the four datasets.
Movie-Ticket Booking, Restaurant Reservation, and Taxi Ordering are three single-domain datasets \cite{ li2016user, li2018microsoft} whose goal is to build a dialog system that helps users find information about movies, restaurants, and taxis and book their tickets. Multiwoz  \cite{budzianowski2018multiwoz} is a multi-domain, multi-content corpus containing seven domains, including attraction, hospital, policy, hotel, restaurant, taxi, and train.

\begin{table}[H]
	\resizebox{\columnwidth}{!}{
		\begin{tabular}{ccccccc} 
			\hline 
			Task&Intents&Slots&Dialogues&Domain\\
			\hline  
			Movie-Ticket Booking&11&29&2890&1\\
			Restaurant Reservation&11&30&4103&1\\
			Taxi Ordering&11&29&3094&1\\
			MultiWoz&13&25&8438&7\\
			\hline
		\end{tabular}
	}
	\centering
	\caption{Comparison of four datasets, Movie-Ticket Booking, Restaurant Reservation, Taxi Ordering, and Multiwoz.}
	\label{tab:dataset}
\end{table}

For each conversation, the user simulator randomly samples a user goal to interact with an agent to complete this user goal. In Multiwoz, the user is allowed to change his goal during the session, and each dialogue can contain more than one domain. Therefore, the Multiwoz dataset is inherently challenging due to its multi-domain setting and multi-intent setting.

Besides that, for Restaurant Reservation, and Taxi Ordering datasets, we performed database cleansing. This was due to the fact that they had too many user goals that were inherently unsuccessful, i.e., there were no entries in the database that satisfied that user goal causing that user goal to never succeed. Therefore, we eliminated those data and replaced them with new user goals and corresponding entries in the database to better display the experimental results.

\end{document}